# Effect of solute atoms segregation on Al grain boundary properties by First-principles study

Xuan Zhang [a], Liang Zhang [a,b,*], Zhihui Zhang [a], Xiaoxu Huang [a,b]

[a] *International Joint Laboratory for Light Alloys (MOE), College of Materials Science and Engineering, Chongqing University, Chongqing 400044, China*

[b] *Shenyang National Laboratory for Materials Science, Chongqing University, Chongqing 400044, China*

**Abstract:** First-principles calculations were carried out to study the segregation behavior of Mg, and Cu and their effect on the energy and mechanical properties of different Al grain boundaries (GBs). Four symmetrical tilt GBs were selected for study, namely Σ5[001](210) GB, Σ5[001](310) GB, Σ9[110](221) GB, and Σ11[110](332) GB. The results show that both Mg and Cu have a segregation tendency at the GBs, and the segregation tendency of Cu is stronger than Mg. Mg is prone to form substitutional segregation at the GBs, but Cu is more likely to segregate at the interstitial sites. The segregation of Mg and Cu can reduce GB energy, and the GB energy continues to decrease with the increase of the segregation concentration. First-principles calculation tensile test shows that the segregation of Mg has a negative effect on the strength of GBs, and the GB strength decreases with the increase of the Mg concentration, while the GB strength was gradually enhanced with the increase of the Cu concentration. The strength of Σ5(210) GB and Σ9(221) GB are more sensitive to the segregation of solute atoms than the other two GBs. By calculating the charge density and the density of states of the pristine and the segregated GBs, it was found that the segregation of Mg caused charge depletion and structure expansion at the GBs, while the segregation of Cu increases the charge density of GBs and form new bonds with the surrounding Al atoms. The results provide useful information for improving the mechanical properties of materials by using the concept of GB segregation engineering.

___________________________

[*]Corresponding author. Email: liangz@cqu.edu.cn (L.Z.)

# 1. Introduction

Grain boundaries (GBs) are important microstructures of metallic materials, which play a significant role in the physical and mechanical properties of metals. By designing and changing the GB characteristics, properties of metal, such as toughness, tensile strength, conductivity, and corrosion resistance could be improved substantially [1-8]. GB segregation is the tendency of solute atoms to spontaneously concentrate towards GB [9], which is an effective way to change the properties of GBs, including cohesive energy, mobility, transport coefficients, nucleation of dislocations, etc. Solute segregation at GBs can enhance and improve the stability and strength of materials [3, 10-13]. For example, Li et al. [14] prepared nanostructure Fe by severe plastic deformation with tensile strength up to 7 GPa, while the high strength of Fe was attributed to the segregation of the C in GBs. Likely, an ultra-refined age-hardened Al-Cu-Mg alloy obtained by the high-pressure torsion method has a tensile strength of about 1 GPa. The structural characterization shows that the segregation of Mg and Cu plays an important role in the high strength of the alloy [15]. Detor et al. [16, 17] found that nanocrystalline Ni with W segregation fabricated by the electroplated method only exhibited small grain coarsening, with the average grain size increasing from 20 nm to 28 nm. In contrast, the grain size of pure Ni is significantly coarsened up to 200 nm, indicating that the stability of Ni GB with W segregation has been improved obviously. Similarly, in the ultrafine lamellar Al-0.3%Cu alloy produced by high-strain deformation, the segregation of Cu at the Y-junction of GBs reduced the energy and the mobility of lamellar boundaries and Y-junctions, which increase the stability and inhibited the GB coarsening during recovery annealing [18]. The above experimental results indicate that GB segregation is a promising method to simultaneously improve the strength and stability of materials.

Although some information on GB segregation can be obtained using characterization techniques, some important physical phenomena in picosecond and femtosecond time domains are difficult to be studied experimentally, such as the evolution of GB structure with solute atom segregation and the effect of solute atom segregation on mechanical behaviors of GBs. These problems, which associate with the accuracy and detail of the atomic scale of GBs, have hindered the quantitative study of GB segregation by only characterization techniques. With the development of high-performance computing, computational simulation has become an effective method to investigate GB properties [19-21]. First-principle calculation based on density function theory (DFT) can help acquire not only the basic energy information of GB and details of solute segregation behavior but also the strengthening mechanism of solute by calculating the electronic structure. This method has been widely used to investigate solute segregation behavior and its effect on GB stability and strength [22-24]. For example, by using first-principle calculations, Bauer et al. [25] found that the energy of Σ5(210) GB in Fe decreases with the increase of Zn segregation. Huang et al. [26] reported that solutes with larger atomic radius were more prone to form segregation, and improve the stability of Σ5(310) GB in Cu. In addition, solute segregation could influence the structure and strength of GB. It was found that Sc atoms segregated at Al Σ5(210) GB would destroy the symmetric structure of GB, but it strengthens the GB by increasing the charge density of GB [27]. Research on H embrittlement at GBs by the DFT method has made some progress. Luo et al. [28] reported that the intergranular strength of Ti was reduced with H segregation. The weakening effect is the result of the creation of H–Ti bonds, which are weak than the existing Ti–Ti bonds. Schuler and colleagues [29] investigated the co-segregation of H, P at α-Fe GB and the corresponding embrittlement effect. It was found that P-H interactions lead to increased P segregation at GBs and cause additional embrittlement of GBs compared to the case where P and H are considered separately.

Both Cu and Mg are the main alloying components of Al alloys, which have strong solution-strengthening effects and could form precipitation. In addition, previous research works have shown that Cu and Mg have a strong tendency to segregate at the Al GBs [18, 30-32]. For example, 3DAP characterization shows the abundant segregation of Mg and Cu at GBs in an ultrafine-grained Al-Zn-Mg-Cu alloy [30]. First principle simulations were carried out on the effect of Mg solute atoms on the properties of Al GBs. For example, Liu et al. [33] reported that the Mg segregation results in a decrease in the strength of Al Σ11(113) GB. Similarly, Zhao et al. [32] and Hu et al. [34] reported that the strength of Al Σ5(210) GB was decreased by the Mg atom. According to the above studies, the main reason for the decrease in GB strength is the charge depletion caused by the segregation of solute atoms at GB. In addition, Razumovskiy et al. [35] and Zhang et al. [36] found that Mg segregation plays a little role in the strength of Σ5(210) GB in Al. They demonstrated that the charge transform effect of the Mg atom offsets the charge depletion effect caused by Mg. Therefore, it seems that there is a controversy on the effect of Mg on the strength of different Al GBs. On the other hand, simulations suggest that the segregation of Cu can improve the strength of GB, but most of the existing computational studies are limited to a specific GB. For example, Zhao et al. [32] found that Cu would increase the strength of Al Σ5(210) GB by forming interstitial segregation. Hu et al. [34] predicted that Cu strengthen the Al Σ5(210) GB when Cu formed substitutional segregation at GB. In what form Cu segregates in the different GBs, and how it contributes to the strength of the various GBs are still not clear. Therefore, a comprehensive study is necessary to further understand the segregation behaviors of solute atoms at Al GBs.

In this work, first-principle calculations were carried out to investigate the segregation of Mg and Cu solute atoms on different GBs of Al, and the effect of segregation on the energy and mechanical properties of the GBs was studied. In the first part, the GB energy and segregation energy were calculated, and the most favorable sites of Mg and Cu in different GBs were determined by the segregation energy. Then, the segregation energy of GB with different concentrations of Mg and Cu was calculated, and the effect of solute concentration on the energy features and structural properties of the GBs was investigated. In the second part, we used the first-principle tensile test to investigate the influence of Mg and Cu segregation on the theoretical strength of GBs, and the effect of the concentration of segregation on GB strength was considered. The electronic structure analysis and partial density of state analysis methods were used to explain the underlying mechanisms of how solute atoms affect the structural and mechanical properties of the GBs.

## 2. Methodology

### 2.1 First-principle calculations

The first-principle calculations in the present work were carried out by using the Vienna ab initio simulation package [37, 38] with the projector augmented wave method to describe the electron-ion interactions [39, 40]. The exchange-correlation potential was treated using the generalized gradient approximation (GGA) in the Perdewe Burkee Ernzerh form [41]. Also, for Brillouin zone (BZ), the Monkhorst-Pack scheme [42] in combination with the linear tetrahedron method including Blochl corrections [43] was applied to generate the K-point mesh. A cutoff energy of 350 eV and the k-points sampling of 6×5×3, 6×4×2, 9×3×2, and 7×4×2 for four kinds of Al GBs respectively to ensure that total energy differences were less than 1meV/atom. Convergence criteria for total energy and forces were $10^{-6}$ eV and $10^{-2}$ eV/Å in the structural relaxations process.

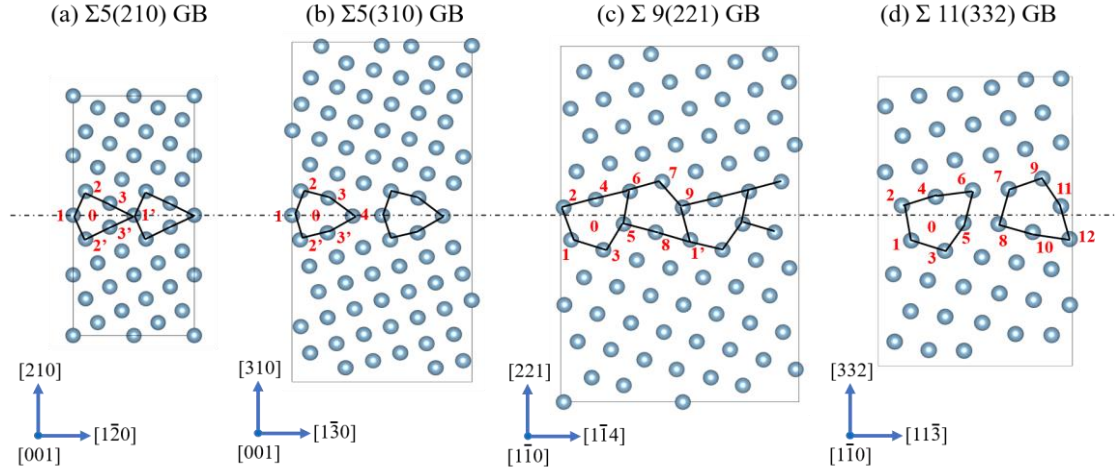

**Figure 1.** Atomic structure of the pristine Al grain boundaries. (a) Σ5[001](210) GB, (b) Σ5[001](310) GB, (c) Σ9[110](221)GB, and (d) Σ11[110](332) GB. The solid line outlines the periodical structural units of the GBs. The number '0' represents the interstitial site and the number '1-12' represent the substitution sites of the structural unit when segregation of solute atoms at GB. The dotted line indicates the symmetry of the GB structure.

**2.2 Grain boundary models**

Four GBs were selected as the representative grain model in this study, namely the Σ5[001] (210) GB, Σ5[001] (310) GB, Σ9[110] (221) GB, and Σ11[110] (332) GB. They are constructed by the coincidence site lattice (CSL) method and the GBs with the most stable structures were chosen as the model of the present work [44-47]. Here, the [001], [110] and (310), (210), (221), (332) denote crystal miller indices of the tilt axis and the habit plane for the GB models, respectively. Fig. 1 illustrates the atomic structure of the pristine grain boundary without segregation. There are 80 and 152, 152, and 92 Al atoms in four supercells respectively. Before building the GB model, the equilibrium lattice constant of Al 4.041Å is calculated with a bulk (3×3×3) supercell model containing 108 atoms, which is agree well with the experiment results of 4.032 Å [48] and the calculation result of 4.046 Å [35]. Taking the Σ5(210) GB as an example, each Σ5(210) supercell contains 20 layers of (210) planes stacking in the perpendicular direction to the GB planes. Due to the periodicity, the supercell contains two GBs located at the middle and bottom of the model as portrayed in Fig. 1. The micro grain is stacked periodically along the direction which is perpendicular to the GB in four GBs and tilted with respect to the [001] and [110] direction. This stacking sequence is reversed after a half period along the direction, forming the mirror symmetry in the CSL GB model. The other three GBs are constructed by the same method. The initial configuration of GB models before atomic relaxation is $(2, \sqrt{2}, \sqrt{5}) a_0$, $(2, 2\sqrt{2}, 4\sqrt{2})a_0$, $(\sqrt{2}, 3\sqrt{2}, 4\sqrt{2})a_0$, $(\sqrt{2}, \sqrt{11}, 5) a_0$, with $a_0$ being the lattice constant of Al.

**2.3 Tensile test calculations**

There are usually two models used in the evaluation of the strengthening and embrittlement effect of solutes in GBs. Griffith model is more appropriate when the crack growth of GB is fast for the diffusion of solute. Rice-Wang model can better depict the situation when GB fracture proceeds slowly enabling the system could reach an equilibrium condition [25]. Mg and Cu solutes are regarded as the immobile solute in Al-Cu-Mg alloys [32]. Therefore, the effect of Mg and Cu segregation on different Al GBs strength are evaluated with a combined methodology of Griffith and first-principle calculation in this study. Two methods are usually used to carry out the first-principle tensile test [29, 49]. The first one is the homogeneous lattice extension (HLE), which imposes the strain throughout the whole lattice homogeneously, taking Poisson's ratio into account, but counting out the fracture plane. The second method is the rigid grain shift (RGS) method, in which the pre-crack plane is set so that the fracture occurs at the GB plane. Both methods have been utilized to investigate the strength property of different GBs, and the result is in good agreement with experiments, including the fracture energy and theoretical strength [32]. Although the Possion effect is not considered in the RGS method, it has little effect on the fracture energy and does not affect the mechanical behavior of different GBs [50]. Therefore, the RGS method was used to perform the first-principle tensile test on GBs in the present work.

# 3. Results and Discussion

**3.1 Segregation site of solute atoms**

The most favorable segregation sites of Cu and Mg along different GBs were analyzed first. According to the previous investigation by Campbell et al. [51] and Liu et al. [52], Cu atoms are more likely to occupy the interstitial hollow site of Σ5(310)GB because of its smaller atomic radius than Al. On the other hand, the substitutional site is more favorable for Mg atoms given their congenial atomic size of Al and Mg. In this study, four GBs are considered, and for each GB, there are several possible segregation positions including substitutional and interstitial sites. Therefore, to determine the GB segregation models with Cu or Mg solutes, the segregation energy ($E_{seg}$) of Mg and Cu at different sites of each GB is calculated according to Eq. 1. There are 10 and 12 substitutional sites in Σ9(221) GB and Σ13(332) GB, but by considering the periodicity, the head site is equal to the tail one, which means there are total 9 sites in Al Σ9(221) GB. Therefore, there are 3, 4, 9, 12 substitutional sites in Σ5(210), Σ5(310) and Σ9(221), Σ11(332) GB, respectively. As shown in Fig. 1, the supercell models containing Mg and Cu residing in different sites of GB have been constructed to calculate the $E_{seg}$ by Eq. 1 [22],

$$E_{seg} = \frac{[(E_{GB}^X - E_{GB}) - N*(E_{bulk}^X - E_{bulk})]}{N} \quad (1)$$

where $E_{GB}^X$ and $E_{GB}$ indicate the total energy of GB with or without Cu, and Mg segregations, and $E_{bulk}^X$ represent the total energy of Al bcc bulk with one Cu or Mg atom dissolved. $E_{bulk}$ indicates the total energy of Al bulk without Mg and Cu solutes. X is the number of Mg or Cu solutes. N is the number of segregation solutes.

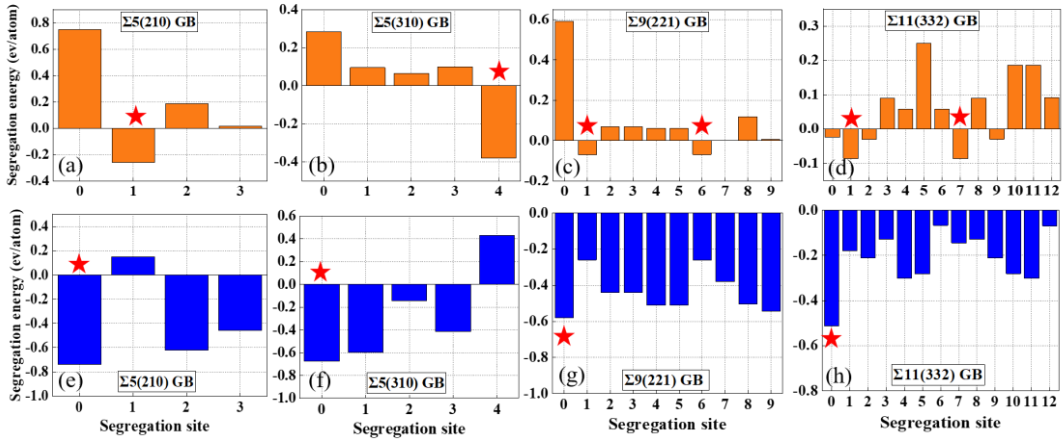

**Figure 2.** The segregation energy of Mg and Cu atoms at different sites of Al GBs. (a)-(d) the segregation energy of Mg, (e)-(h) the segregation energy of Cu. The pentagram indicates the most favorable segregation site of Cu and Mg at the GB.

According to the results of segregation energy, the most favorable segregation site of Cu and Mg at each GB was decided, as indicated by the pentagram in Fig. 2. It is found that the segregation energy of Mg at the interstitial site is higher than that at the substitutional site for all GBs, indicating that Mg prefers to segregate in the substitutional site. For example, Mg atoms prefer to segregate at the No.1 site of Σ5(210) GB and the No.4 site of Σ5(310) GB, the corresponding segregation energy is -0.33 ev/atom and -0.38 ev/atom. The $E_{seg}$ has a minimum value when Mg stays at No.1 and No.6 sites in Σ9(221) GB, so the two sites are chosen to be the substitutional site for Mg in Al Σ9(221) GB. Also, it could be seen that there are two same $E_{seg}$ values of Mg in the No.4 site and No.10 site which is lower than that of other sites in Al Σ11(332) GB. On the other hand, the segregation energy of Cu is minimum when Cu resides in the interstitial site of the four GBs, which means that the hollow site of Al GB is more favorable for Cu to stay, which is agree with the calculation work on Σ5(210) GB [32] and experimental results on Σ5(310) GB[52, 53]. Therefore, according to the results of segregation energy, the GB segregation models with Mg and Cu solutes are constructed respectively, as shown in Fig. 3.

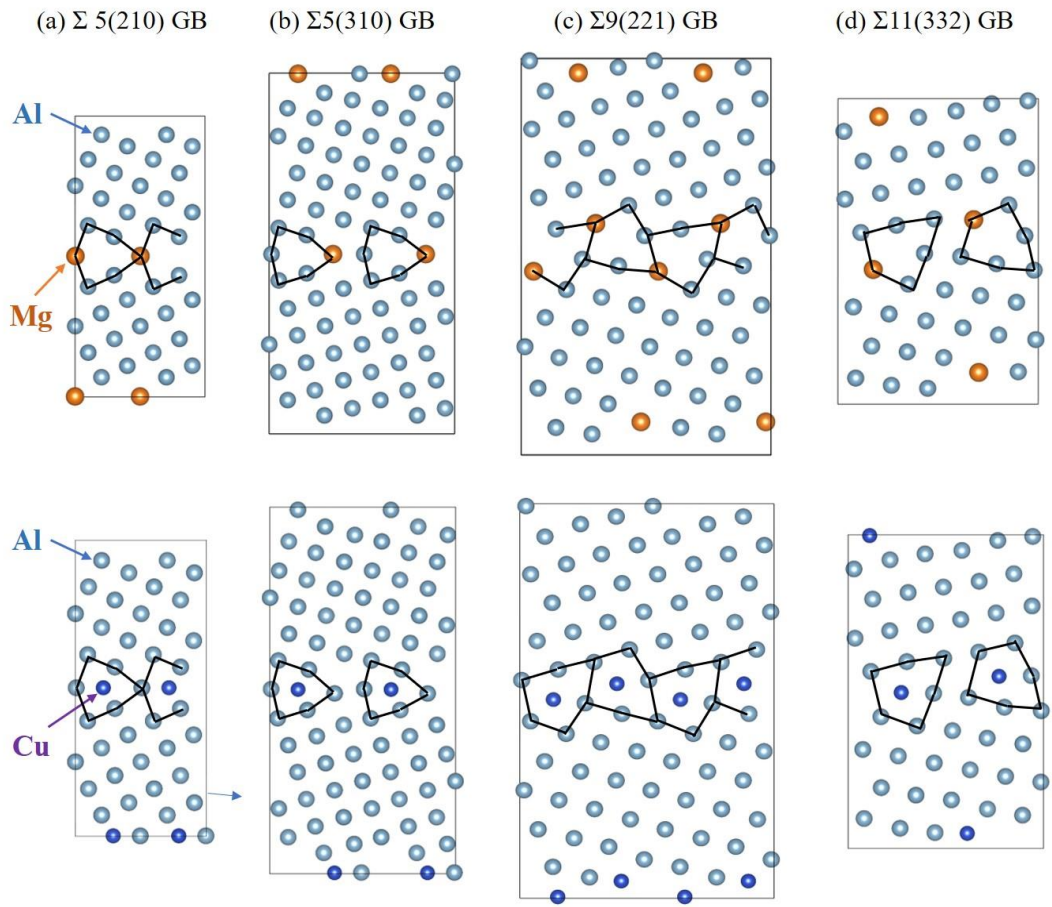

**Figure 3.** Atomic structure of the Al GBs with segregated Mg atoms and Cu atoms. Al atoms are colored gray, Mg atoms are colored orange, and Cu atoms are colored blue. The solid line outlines the periodical structural units of the GBs.

**3.2 GB energy characteristics**

Different concentrations of solute atoms are set up in the segregation models to investigate the influence of the solute concentration on GB energy and GB segregation energy. When all the segregation sites (including substitutional and interstitial sites) at GB are filled up with Mg or Cu atoms, the concentration was defined as one monolayer (1ML). Accordingly, 0.75ML means there are 3/4 sites at GB filled up with the solute atoms, and so on in a similar fashion. The GBs with different concentrations of Mg and Cu solute are constructed, and GB energy and GB segregation energy are calculated. Grain boundary energy ($\gamma_{GB}$) can be obtained by Eq. 2,

$$\gamma_{GB} = \frac{E_{GB} - NE_{Al}}{2S} \qquad (2)$$

Where $E_{GB}$ denotes the total energy of GB supercells, N represents the number of Al atoms, S means the area of the interface, $E_{Al}$ represents the chemical potential per atom, which can be calculated by Al fcc supercell, the 1/2 in Eq. 2 is due to there are two GBs in the supercell. The calculated GB energies of Σ5(210) GB, Σ5(310) GB, Σ9(221) GB, and Σ11(332) GB are 0.526, 0.477, 0.476, and 0.425 J/m$^2$, respectively.

The $\gamma_{GB}$ of the segregated GBs with different Mg or Cu concentrations are calculated by Eq. 3,

$$\gamma_{GB} = \frac{E_{GB}^{Nx} - N_{Al}E_{Al} - N_x E_x}{2S} \qquad (3)$$

Where $E_{GB}^{Nx}$ is the total energy of different GBs containing N Mg or Cu atoms, $N_x$ and $N_{Al}$ is the number of Mg, Cu and Al atoms, and x is the Mg or Cu solutes. $E_x$ represents the energy of a single Mg or Cu atom which can be obtained by the calculation of Mg and Cu bulk, respectively, and S is the area of the GB interface. Here the scaling factor of 1/2 means there is two GBs in the simulation supercell.

Fig. 4 (a) and (b) show the calculated results of GB energy as a function of different concentrations of Mg and Cu solutes. It is found that the segregation of Mg and Cu can decrease the GB energy of the four GBs, and with a higher concentration of the Mg or Cu solutes, GB energy decreases furtherly. 1ML Mg residing the Σ11(332) GB and Σ9(221) GB can result in an almost 20% reduction of GB energy, while the same concentration of Mg caused an almost 40% decrease of GB energy in the cases of Σ5(210) GB and Σ5(310) GB. In addition, Cu segregation at GBs has a more significant effect on the reduction of GB energy than Mg. For example, 1ML Cu would result in a 90% and 97% decrease in GB energy for Σ5(310) GB and Σ11(332) GB, respectively. In particular, 1ML Cu residing in the Σ5(210), Σ9(221) GB brings the GB energy down to negative values. GB energy could be negative means that GBs are in a condition that is more stable than the Al bulk without GB. This situation could happen in nanostructure metals. Because of the high density of GB, much more solutes could segregate at the GBs and increase the stability of GB. The reason that GB energy decreases with the increase of solute segregation concentration can ascribe to the release of the strain energy at the GB [54, 55]. Since the Cu segregation at the hollow site could better decrease the strain energy than Mg, the GB energy with Cu segregation is much lower than that of Mg. The results of GB energy have an implication that the segregation of Cu and Mg can effectively decrease the GB energy of Al and improve the stability of Al nanocrystals.

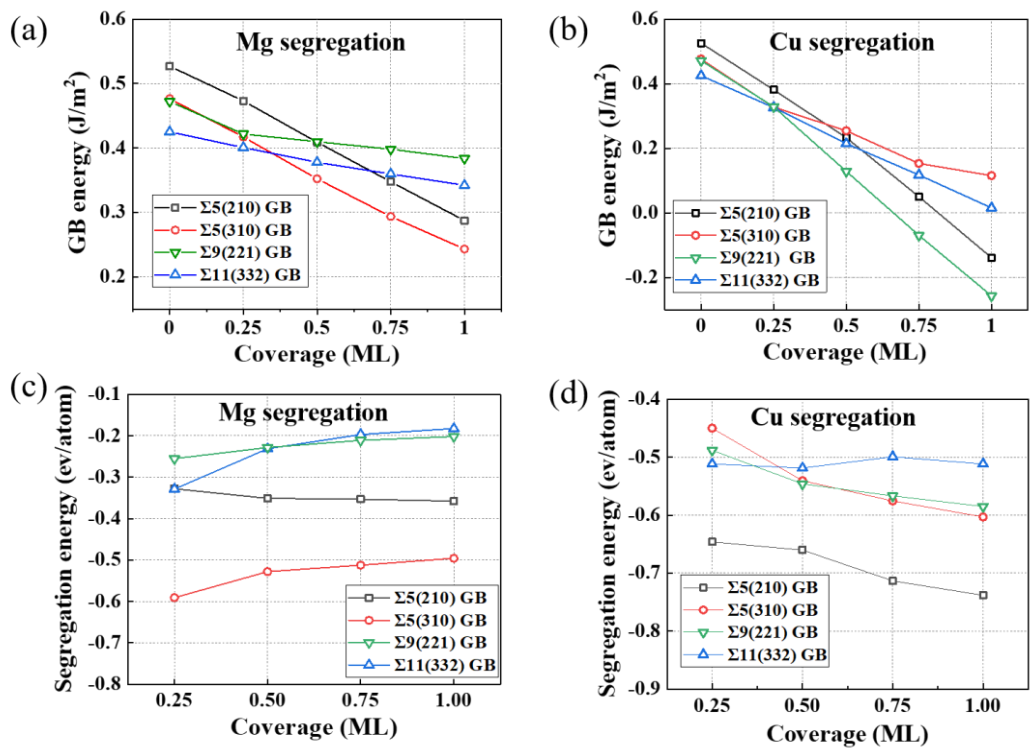

**Figure 4.** (a) and (b) GB energy of Σ5(210) GB, Σ5(310) GB, Σ9(221) GB, Σ11(332) GB with different Mg, Cu segregation concentration, (c) and (d) Segregation energy of Mg and Cu at different GBs.

Fig. 4(c) and (d) show the segregation energy of the GBs with different concentrations of Mg and Cu solutes. It is worth noting that the segregation energy remains negative value in all simulation cases, indicating that both Mg and Cu have tendency to segregate at the GBs. In addition, the segregation energy of Cu at different GBs is lower than that of Mg, indicating that Cu has a stronger motivation to migrate from the inside grains to GBs than Mg. The segregation energy of Cu at Σ5(210) GB is lower than that of other GBs, which means that Cu prefers to segregate to Al Σ5(210) GB among the four tested GBs. On the other hand, Mg prefers to stay at Σ5(310) GB throughout the concentration range for the lowest segregation energy than the value of other GBs. The segregation energy of Cu at Σ11(332), Σ5(210), and Σ5(310) GBs decrease with the increase of Cu concentration, but there is no evident relationship between the concentration and segregation energy at Σ9(221) GB, as shown in Fig. 4(d).

Contrary to the cases of Cu segregation at GBs, the segregation energy of Mg at Σ5(310), Σ9(221), and Σ11(332) GBs increases slowly with the increase of Mg residing at the GBs. The segregation energy of Mg segregation is overall higher than the value of Cu, which means that Cu atoms have a stronger tendency to segregate at Al GBs than Mg. To explain the reasons that caused this difference, Bader charge calculation was carried out for the solute of Mg and Cu at different GBs. The Bader charge method is a useful method to calculate the electronic charge associated with an atom [22]. By this method, the change of Mg and Cu valence electrons after GB segregation can be obtained. Because of the higher electronegativity than the Al atom, the Cu atom obtains electrons from the Al atom, the higher value of Bader charge means a stronger charge transfer effect, while the Mg atom donates electrons to the Al atom for lower electronegativity, and a higher value means weaker charge transfer. The higher value of charge transfer means a stronger electronic interaction between the atoms. Table 1 lists the Bader charge value of Mg and Cu at different GBs. It is found that the Bader charge of Mg atoms in all GBs is higher than that of Mg atoms at the Al bulk (0.56 e), indicating that the Mg atom has a weaker electronic interaction with the Al atom at GBs than in the bulk. On the other hand, the Bader charge of Cu atoms in all GBs is higher than that of Cu atoms at the Al bulk (11.7 e), which means a stronger electronic interaction between Cu and Al atoms at GBs than in the bulk. The stronger electronic interaction between Cu and Al atoms leads to the higher tendency of Cu atoms to segregate at Al GB than Mg atoms.

**Table 1.** Bader charge value of Mg and Cu at Al bulk and different Al GBs, the unit is e.

| Solute | Al bulk | Σ5(210) | Σ5(310) | Σ9(221) | Σ11(332) |
|---|---|---|---|---|---|
| 1.00 ML-Mg | 0.56 | 0.61 | 0.67 | 0.70 | 0.69 |
| 0.25 ML-Mg |  | 0.60 | 0.57 | 0.60 | 0.58 |
| 1.00 ML-Cu | 11.7 | 13.08 | 13.12 | 13.32 | 13.22 |
| 0.25 ML-Cu |  | 13.15 | 13.37 | 12.78 | 12.60 |

**3.3 GB mechanical properties**

To clarify the relationship between GB energy characteristics and mechanical properties, first-principle tensile test calculations are performed in this part to evaluate the influence of Mg and Cu segregation on the theoretical strength of different GBs.

The tensile tests are implemented after the relaxation of the GBs model with or without solute segregation. During the test, the GB model is rigidly separated with a certain displacement (from 0.2, 0.4, 0.6, 0.8 Å increased to 6.0 Å). The separation is perpendicular to the direction of the GB plane. For each displacement, two kinds of calculation are performed, (i) RGS without atomic relaxation, (ii) RGS with full atomic relaxation.

Then, the separation energy of GB model with a certain displacement can be calculated by Eq. 4 [56],

$$E_{sep} = \frac{E_{sep}^{x} - E_{GB-y}}{2S} \quad (4)$$

Where $E_{sep}$ denotes the separation energy, while the $E_{GB-y}$ represents the energy of the pristine GB model without or with y segregation, (y means Mg or Cu). $E_{sep}^{x}$ represents the energy of the GB model with the separation distance of x, and S represents the area of the fracture GB. Through the equation proposed by Rose et al. [57], the relationships between separation distance and separation energy can be fitted by Eq. 5 and Eq. 6,

$$f(x) = E_{frac} - E_{frac}\left(1 + \frac{x}{\lambda}\right)e^{\left(-\frac{x}{\lambda}\right)} \quad (5)$$

$$E_{frac} = \frac{E_{sep}^{\infty} - E_{GB}}{2S} \quad (6)$$

Here, λ is the characteristic distance [58] and $E_{frac}$ is fracture energy. $E_{sep}^{\infty}$ is the total energy of the GB model when the separation distance is so far that the two parts of the GB break. After that, the stress-strain curve can be calculated by the derivative of the Eq. 5 as Eq. 7,

$$f(x)' = \frac{E_{frac}x}{\lambda^2}e^{-\frac{x}{\lambda}} \quad (7)$$

When the separation distance x is equal to λ, the maximum strength also can be obtained by:

$$ð_{max} = f(x)' = \frac{E_{frac}}{\lambda e} \quad (8)$$

the separation energy and theoretical strength as a function of separation distance (before and after adding different concentrations of Mg or Cu into the four kinds of GBs) are calculated by Rose equation [57].

Fig. 5 shows the mechanical responses of Al Σ5(210) GB with different concentrations of Cu and Mg solutes without atomic relaxation. The results of bulk Al with (210) plane and pristine Al Σ5(210) GB are also plotted for comparison. As can be seen from Fig. 5 (a), the separation energy rises rapidly with the increase of separation distance and finally reaches a constant value. The separation energy of bulk Al is calculated as 2.14 J/m$^2$, which agrees well with the previous simulation results (2.21 J/m$^2$) [34] as well as the experimental results of 2.30 J/m$^2$ [59]. The separation energy of pristine Σ5(210) GB is calculated as 1.80 J/m$^2$, which is lower than that of bulk Al, implying that the existence of GB weakens the mechanical strength of the matrix. The calculated value is close to the experimental result (1.92 J/m$^2$). The separation energy of GB with 0.25ML Cu is higher than pristine Al Σ5(210) GB, and with the increase of Cu segregation concentration, the separation energy increases gradually. The solute segregation of 1ML Cu increases the separation energy to 2.15 J/m$^2$, which is even higher than the value of bulk Al. The result suggests that the segregation of Cu is probable to enhance the tensile strength of Al Σ5(210) GB. On the contrary, in all simulation cases, the separation energy of GB with Mg segregation is lower than the pristine Al Σ5(210) GB, as shown in Fig. 5(d), indicating that the Mg segregation may degrade the cohesion strength of Al Σ5(210) GB. With the increase of Mg segregation, the separation energy decreased further. The segregation of 1ML Mg reduced the separation energy to 1.72 J/m$^2$.

Fig. 5(b) and (e) plotted the evolution of tensile stress as a function of the separation distance with Cu and Mg segregation, respectively. In all simulation cases, the tensile stress climbs rapidly as the separation distance increases, after reaching the peak value, the stress drops gradually and reaches zero when the separation distance is about 5Å. The maximum tensile stress corresponds to the theoretical strength of the GB. Fig. 5(c) and (f) plot the theoretical strength of the simulation cases in Fig. 5 (b) and (e), respectively. Compared to the bulk Al, the reduction of the tensile strength is found in the case of Al Σ5(210) GB, from 12.00 GPa to 10.70 GPa. It is found that the theoretical strength of Σ5(210) GB increases with more segregation of Cu at the GB. In particular, when 0.75ML and 1ML Cu reside in Σ5(210) GB, the GB strength rises to 12.81 GPa and 13.44 GPa respectively, which are higher than the strength of bulk Al. On the contrary, the GB strength decreases with incremental Mg segregation. When 1ML Mg segregates in Σ5(210) GB, the strength of GB decreased to 8.95 GPa, a 15.8% reduction compared to the strength of pristine GB, indicating that the segregation of Mg would weaken the strength of Al Σ5(210) GB.

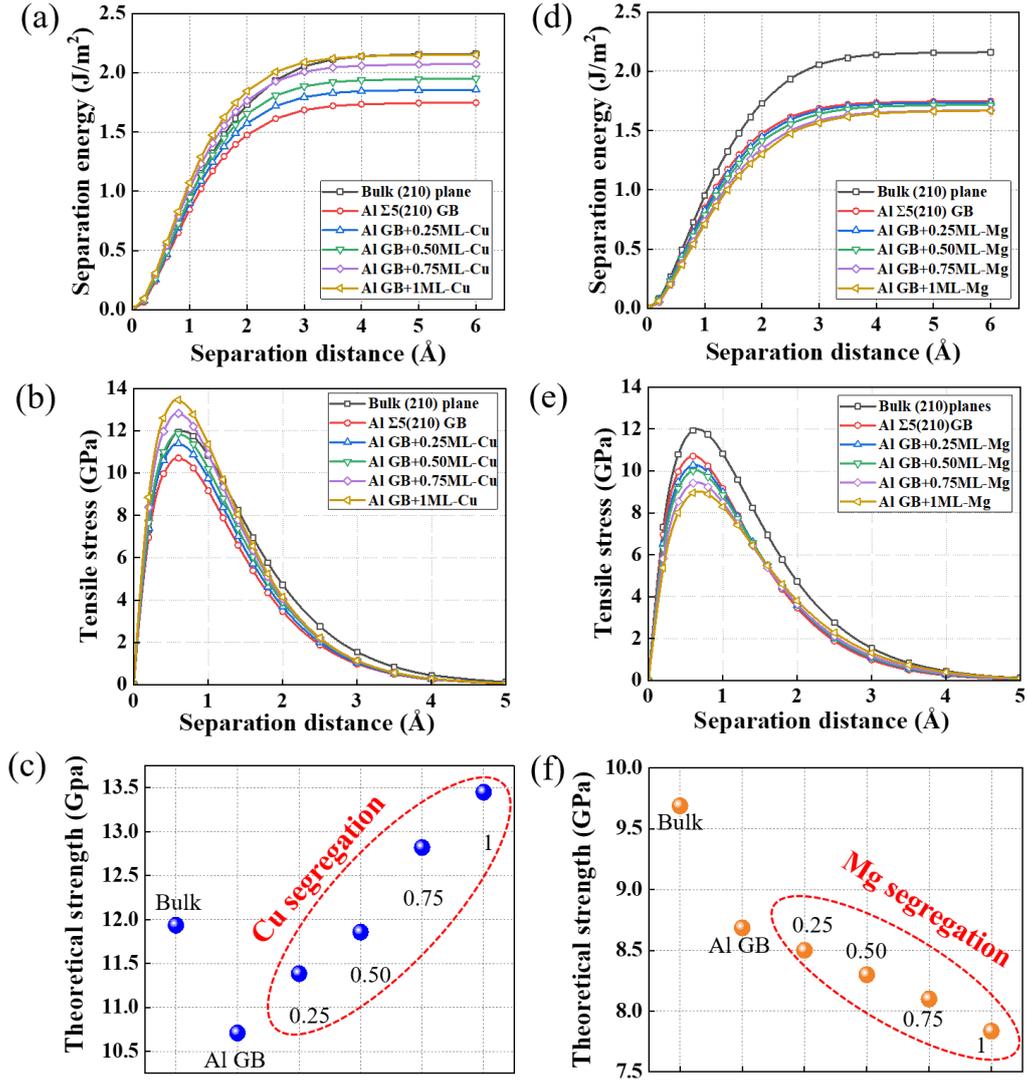

**Figure 5.** Separation energy and tensile stress of Σ5(210) GB with different segregation concentration of Cu, Mg solute atoms.

The mechanical responses of Al Σ5(210) GB with different concentrations of Cu and Mg solutes by full atomic relaxation method were also investigated, as shown in Fig. S1 in the *supplementary materials*. The results of the two RGS methods are collected in Table S1 for comparison. It was found that the structural relaxation decreases the separation energy and tensile strength of the simulation sample in all cases, but the general trend of the results is consistent with that by the non-relaxation method. For example, the separation energy and tensile strength of the Σ5(210) GB still increases with the increase of Cu segregation, and vice versa for the Mg segregation. Since the full atomic relaxation method consumed much more computing time and resources than the non-relaxation method, all calculations were performed without atomic relaxation in the following work.

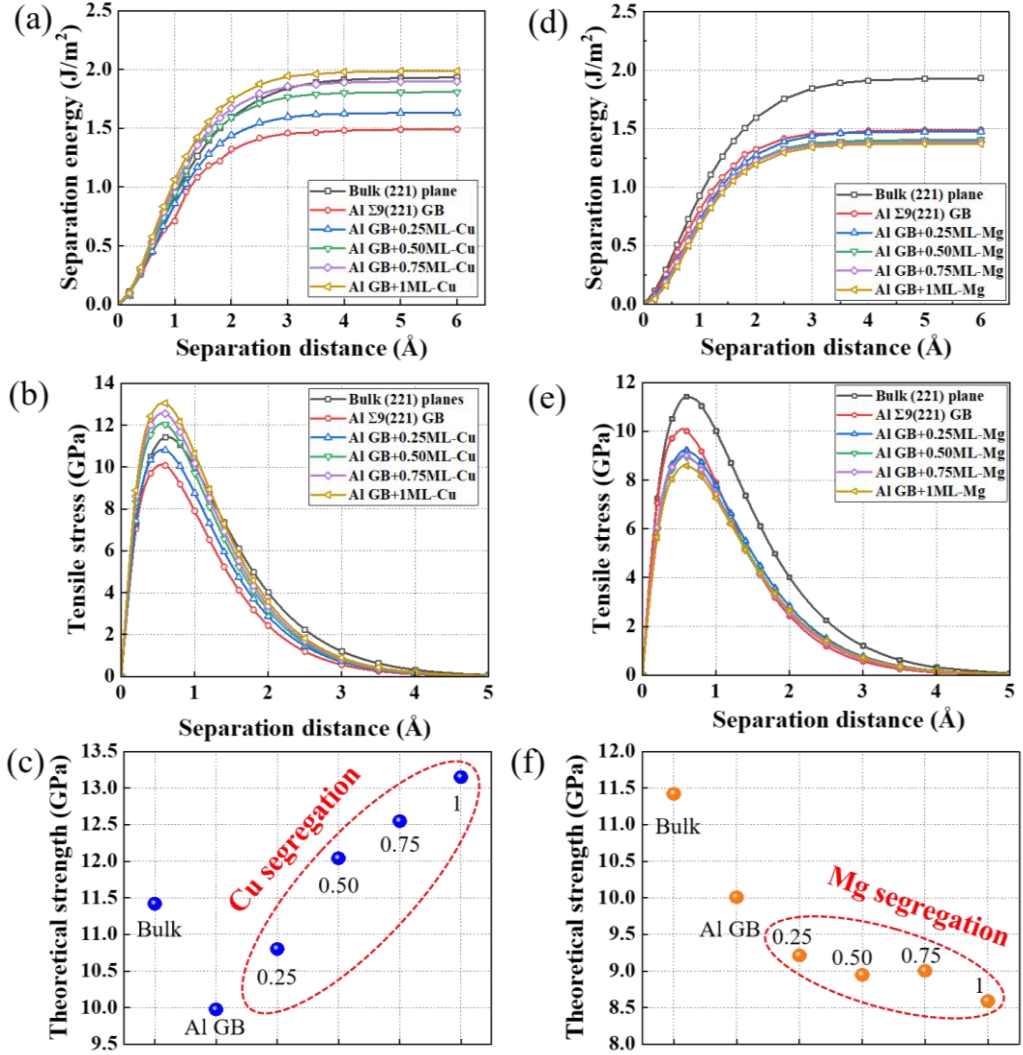

**Figure 6.** Separation energy and tensile stress of Σ9(221) GB with different segregation concentration of Cu, Mg solute atoms.

Fig. 6 shows the separation energy and the tensile strength of Al Σ9(221) GB with Mg and Cu solutes as a function of separation distance. Similar to the case of Σ5(210) GB, the segregation of Cu strengthens the cohesion of Σ9(221) GB, and the separation energy and tensile strength of the GB increase with the increase of Cu concentration. There is a maximum of strength (13.04 GPa) when 1ML Cu segregated at Σ9(221) GB, which is higher than that of pristine GB (10.07 GPa) and the bulk Al with (221) plane (11.41 GPa). The segregation of Mg has a weakening effect on the separation energy and tensile strength of Al Σ9(221) GB, and this trend becomes more obvious with the increase of Mg concentration. The GB strength decreases to 8.64 GPa with 1ML Mg segregation at the GB. It should be noted that the segregation of Mg results in a less decrease in the strength of Σ9(221) GB (12.9%) than that of Σ5(210) GB (15.8%).

The separation energy and tensile responses of Σ5(310) GB and Σ11(332) GB with different concentrations of Cu, and Mg segregation are shown in Fig. S2 and S3, and Fig. 7 plots the theoretical strength of the two GBs. In Fig. 7(a) and (b), the theoretical strength of the two GBs increases with the increase of Cu segregation. The maximum strength (11.62 GPa) of Σ11(332) GB is achieved with the segregation of 1ML Cu, which is higher than the strength of the bulk Al (11.01 GPa). However, with the same concentration of Cu segregation, the strength of Σ5(310) GB is at a similar level to that of the bulk Al. In Fig. 7(c) and (d), the GB strength with Mg segregation is lower than that of pristine GB for both Σ5(310) GB and Σ11(332) GB, and the value decreases continuously with the increase of Mg concentration.

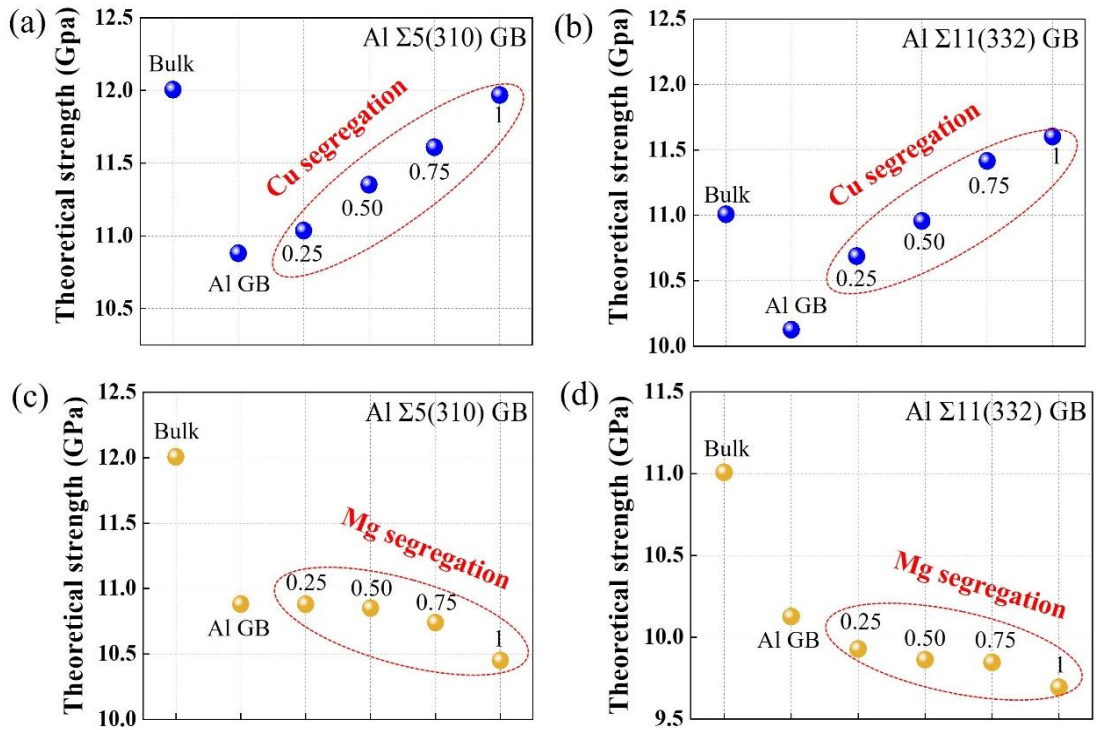

**Figure 7.** The theoretical strength of the (310) plane, (332) plane, and Al Σ5(310)GB, Σ11(332) GB with different Mg and Cu segregation.

For the investigated four grain boundaries, the strengthening or weakening effects of Cu, Mg solute atoms are different. Fig.8 shows the decrease and increase of the GB strength of the four GBs with different concentrations of Cu, Mg solute atoms. It was found that the strength of Σ5(210) GB and Σ9(221) GB decreased more significantly than that of the other two GBs with different Mg concentrations, as shown in Fig. 8(a). When 1ML Mg was segregated at the GBs, the strength of Σ5(210) GB and Σ9(221)

GB decreased by 16.4% and 13.9%, respectively, compared to their corresponding pristine GBs, while the strength of Σ5(310) GB and Σ11(332) GB decreased by 3.9% and 4.2%. The similar situation was found in the case of Cu solute atom, the increases strength of Σ5(210) GB and Σ9(221) GB are more obvious than those of the other two GBs under all the segregation concentrations. When 1ML of Cu was segregated at the GBs, the strength of Σ5(210) GB and Σ9(221) GB increased by 25.5% and 31.8%, respectively, while the strength of Σ5(310) GB and Σ11(332) GB increased by 9.9% and 14.5%. The statistic results lead to two conclusions. On the one hand, the variation of GB strength caused by segregation of the same solute atom at different GBs is different. The strength of Σ5(210) GB and Σ9(221) GB are more sensitive to the segregation of solute atoms than the other two GBs. On the other hand, the influence of Cu on GB strength is stronger than Mg, which is consistent with the influence of both elements on GB energy.

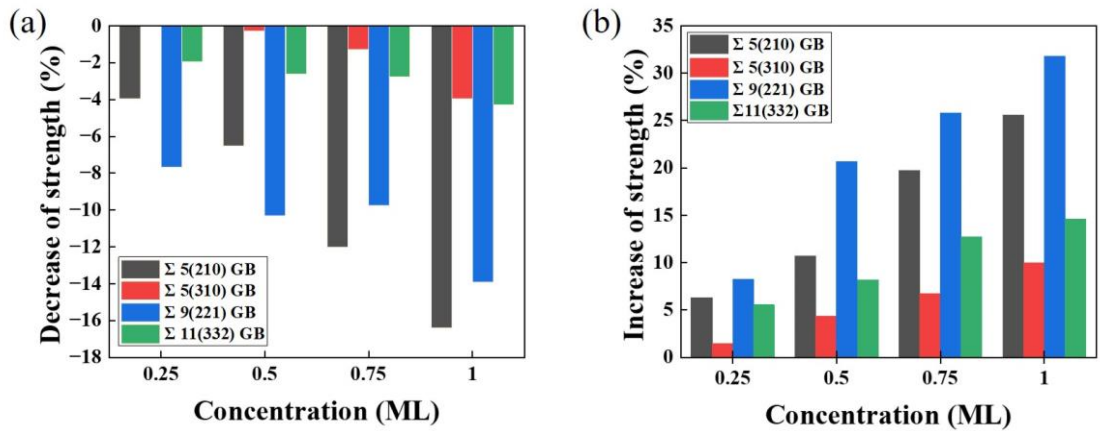

**Figure 8.** (a) The decrease of strength caused by Mg segregation, (b) the increase of strength caused by Cu segregation.

### 3.4 Charge density analysis

The above results show that the segregation of Cu, Mg solute atoms can have a strong effect on the structure and energy of GBs, and the two elements have completely different effects on the theoretical strength of GBs. The segregation of Mg has a negative effect on the strength of GB, while the segregation of Cu is conducive to improving the strength of GB. It is necessary to further explore the relationship between solute elements, GB structure, and GB strength.

The distributions of charge density can be used to describe the chemical bond property between the atoms. The charge density depletion and structure expansion are generally considered negative effects leading to the reduction of GB strength. Therefore, to understand the strengthening or weakening effect of Mg and Cu in GBs, the charge density of the GBs is calculated under tensile deformation. Fig. 9 shows the charge density distributions of Σ5(210) GB at different separation distances. For the pristine Al Σ5(210) GB, it can be found that there is strong charge accumulation between Al (4) atom and Al (4') atom, indicating that the Al (4)-Al (4') bond is the main contribution to the GB strength. Also, there is a low charge density region between Al (1-2-3-3'-2') atoms due to the free volume at the boundary, as shown in Fig. 9(a). With the increase of the separation distance, the charge density around the boundary plane decreases gradually. When the separation distance increases to 0.6 Å, the charge density between Al (4) and Al (4') atoms is about 0.024 e/bohr$^3$. At the separation distance of 1.4 Å, the low charge density region almost extends to the entire boundary plane, and the charge density between Al (4) and Al (4') atoms reduced significantly compared to the original GB plane without separation. The GB is separated when the distance increases to 1.6 Å and the charge density at the boundary plane drops to zero.

Fig. 9(b) shows the variation of charge density of Σ5(210) GB with Mg segregation at different separation distances. Compared to the pristine GB, the participation of Mg caused the expansion of Σ5(210) GB. For example, as shown in Fig. 9(b1), the distance between Al (3) and Al (3') atoms increases from 4.02 to 4.26 Å, and the distance of the Al (4)-Al (4') bond that mainly contributes the strength of GB has increased from 2.58 to 2.67 Å. As a result, the expansion of GBs caused by the Mg atom makes the charge density between Al (4)-Al (4') region down from 0.042 to 0.039 e/bohr$^3$. As shown in Fig. 9(b2), the charge density at the boundary plane decreases quickly with the increase of the separation distance. It can be found that the Mg segregation would lead to substantial charge density depletion around the Mg atom. For example, the newly formed Mg-Al (2) bond is weaker than the bond of Al (1)-Al (2) at pristine GB. Therefore, at all separation distances, the charge density at the boundary is lower than the value of the pristine Σ5(210) GB. When the distance increase to 1.4 Å, the low charge density region runs through the entire boundary plane. The intrinsic electron character of Mg that have only two valence electrons results in fewer electron interaction between the Mg and Al [34]. The larger radius of the Mg atom leads to the expansion of Al GBs is another reason for the decrease in strength. Fig. S4 shows the charge density distributions of the pristine and segregated Σ5(210) GB using full atomic relaxation method. The results of both calculation methods show the same trend.

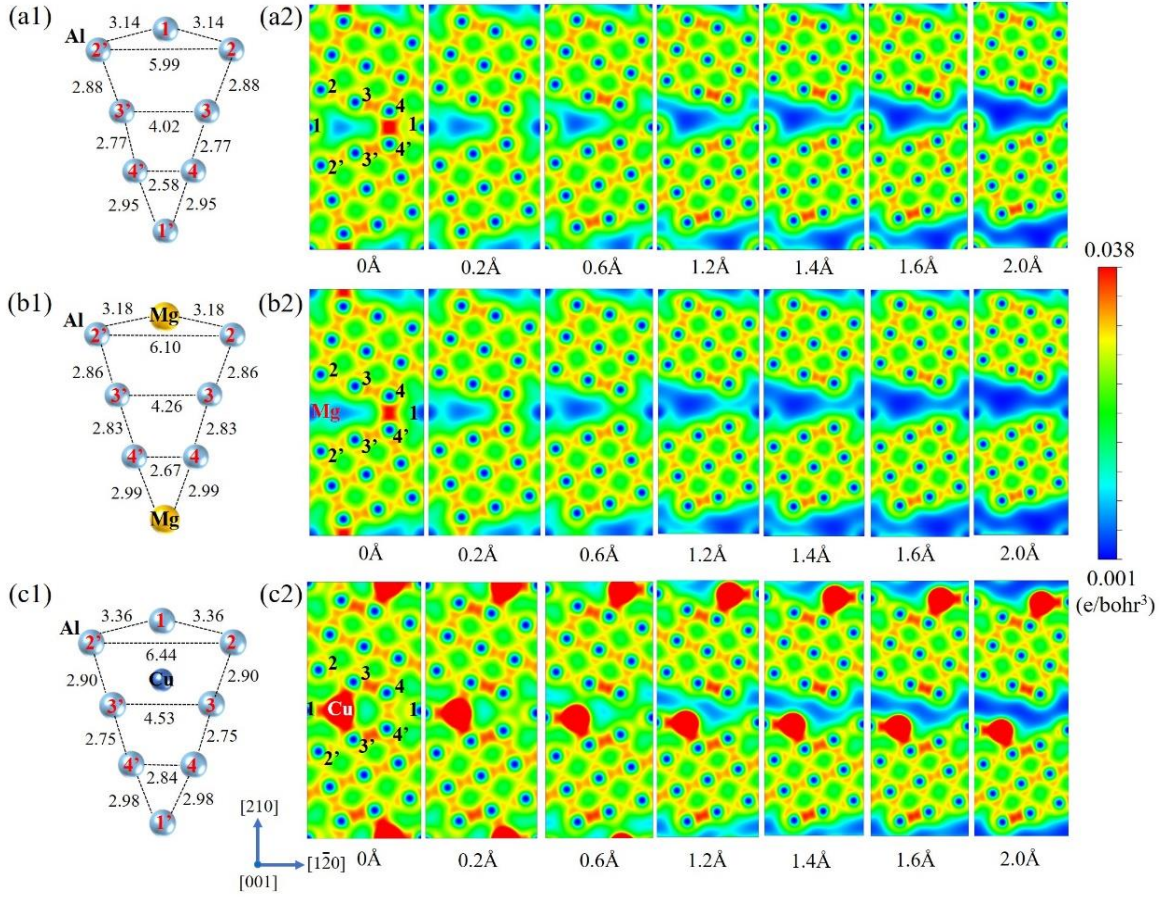

**Figure 9.** (a1)-(c1) Atomic structure of the pristine Al Σ5(210) GB in [001] plane, GB with Mg segregation, and GB with Cu segregation. (a2)-(c2) charge density distributions of the pristine Al Σ5(210) GB, GB with Mg segregation, and GB with Cu segregation at different separation distances.

The distribution of charge densities of Σ5(210) GB with Cu segregation at different separation distances are shown in Fig. 9(c). It can be found that the segregation of the Cu atom extends the GB structure. As shown in Fig. 9(c1), the distance of Al (3)-Al (3') increased from 4.02 to 4.53 Å compared to the pristine GB, and the distance of the Al (4)-Al (4') bond increased remarkably from 2.58 to 2.84 Å. The results indicates that the existence of the Cu atom results in the decrease on the charge density of the main strength bond. Fig. 9(c2) shows that the charge density between Al (4) and Al (4') atoms decreases from 0.042 to 0.030 e/bohr$^3$. In addition, compared to the pristine GB and the GB with Mg segregation, Cu resides at the boundary not only increases the charge density of the free volume region of the GB but also generates new bonds between itself and neighbor Al atoms, namely Cu-Al (3), Cu-Al (3') and Cu-Al (1). The newly formed Cu-Al (3) bond exhibits a binding charge of 0.042 e/bohr$^3$, which is equal to the value of Al (4)-Al (4') in the pristine GB. The results indicate that after the segregation of the Cu atom, the Cu-Al (3), Cu-Al (3'), and Al (4)-Al (4') both contribute to the strength of the Σ5(210) GB. Therefore, the GB strength increases with the increased concentration of Cu atoms at the boundary. With the increase of separation distance, the expansion of the GB structure makes the charge density of GB reduced rapidly. However, at all separation distances, the charge density of the GB is still higher than that of pristine GB and the GB with Mg segregation, indicating increased GB strength with Cu segregation.

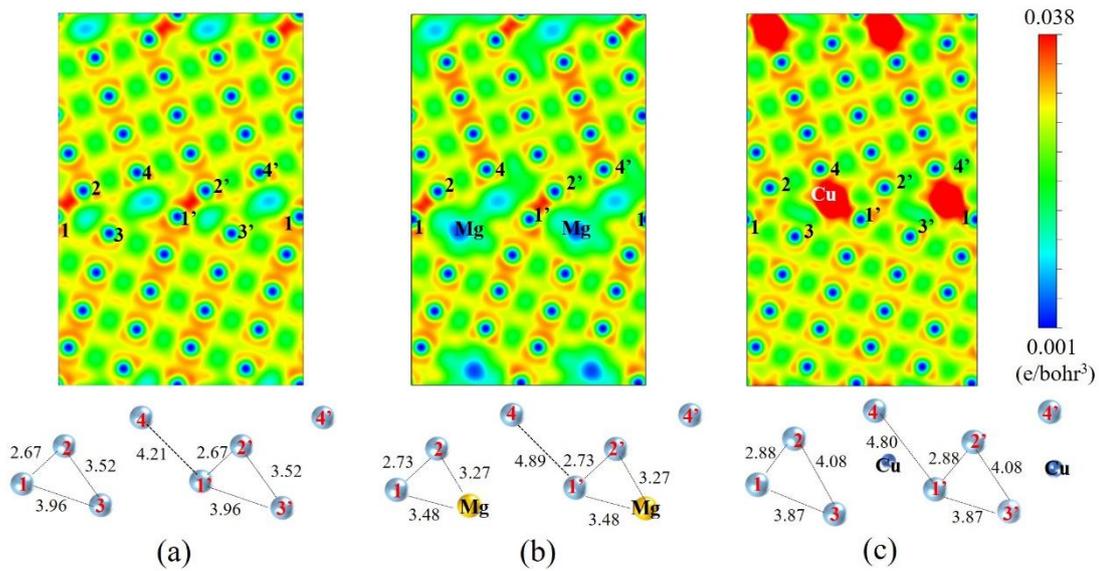

**Figure 10.** Charge density distributions and distance between atoms of Al Σ9(221) GB in [001] plane. (a) Pristine GB, (b) GB with Mg segregation, and (c) GB with Cu segregation.

To understand the effects of Mg and Cu segregations at the Al Σ9(221) GB, the charge density distribution of GB with Cu and Mg atoms is calculated, as shown in Fig. 10. For the pristine Al Σ9(221) GB, Fig. 10(a) shows that Al (1)-Al (2) bond contributes the main strength of GB when fracture happens. As illustrated in Fig. 10(b), the segregation of Mg expands the Al Σ9(221) GB, and the bond length of Al (1)-Al (2) increases from 2.67 to 2.73 Å. Also, the charge density of the Al (1)-Al (2) decreases by 0.001 $e/bohr^3$ for the Mg atom. The decrease of charge density and the expansion of GB structure means that the Mg would cause the weakening of Al Σ9(221) GB. In addition, Mg at Al Σ5(210) GB causes more significant expansion of GB and decreasing in charge density. Hence, the influence caused by the Mg atom at the Al Σ9(221) GB is smaller than that of Mg at Al Σ5(210) GB. On the other hand, the effect of the Cu atom is different from the that of Mg atom at the Al Σ9(221) GB. Fig. 10(c) shows that Cu segregation has strong electron interaction with the surrounding Al atom. Although Cu expands the Al (1)-Al (2) bond from 2.67 to 2.88 Å, the Cu atoms not only increase the charge density of the GB but also form new chemical bonds with the surrounding Al atoms, namely Cu-Al (1) and Cu-Al (4) bonds. The newly formed bond Cu-Al (4) with a charge density of 0.04 e/bohr$^3$, which is higher than that of Al (1)-Al (2), 0.038 e/bohr$^3$. Therefore, the Cu solute is conducive to strengthening the Σ9(221) GB.

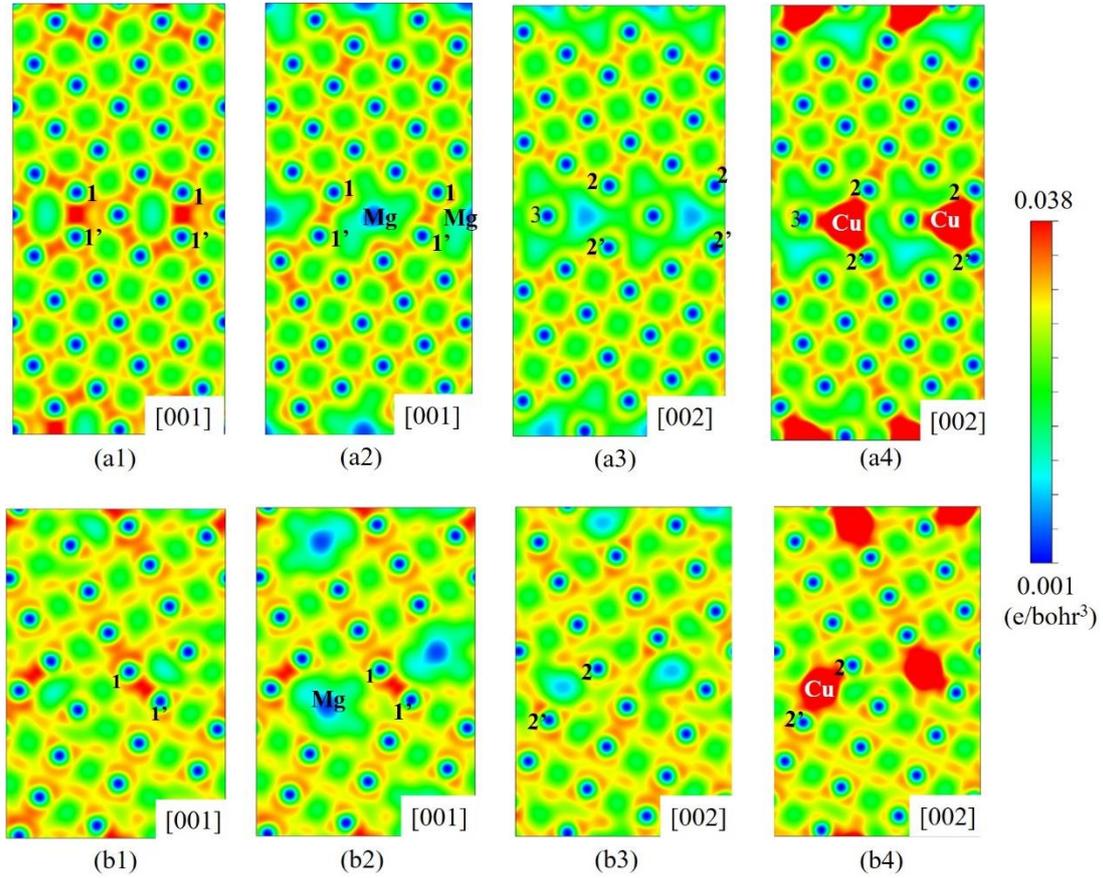

**Figure 11.** (a1)-(a2) Charge density distribution of Σ5(310) GB in [001] plane without and with Mg segregation, (a3)-(a4) Charge distribution of Σ5(310) GB in [002] plane without and with Cu segregation. (b1)-(b2) Charge density distribution of Σ11(332) GB in [001] plane without and with Mg segregation, (b3)-(b4) Charge density distribution of Σ11(332) GB in [001] plane without and with Cu segregation.

Fig. 11 shows the charge density distribution of Al Σ5(310) GB and Σ11(332) GB with Mg and Cu segregation. [001] and [002] represent two successive layers of atomic planes of the simulation model. As shown in Fig. 11(a1) and (a2), compared to pristine Σ5(310) GB, the segregation of the Mg atom decreases the charge density around itself, especially for the Al (1)-Al (1') bond which contributes the main strength of the GB, decrease from 0.041 e/bohr$^3$ to 0.033 e/bohr$^3$. A similar situation is evidenced in the Σ11(332) GB shown in Fig. 11(b1) and (b2), the segregation of Mg caused the charge depletion at the GB plane and increase the low charge density region of the GB. On the other hand, in Fig. 11(a3) and (a4), the segregation of Cu at the [002] plane of Σ5(310) GB decreases the low charge density region Al (2-2'-3) of the pristine GB. Also, the segregation Cu atom forms a new bond with the surrounding Al (2), Al (2'), and Al (3) atoms. The newly formed Cu-Al (2), and Cu-Al (2') bonds have higher charge density (0.042 e/bohr$^3$) than the Al (1)-Al (1') bond (0.04 e/bohr$^3$) which contributes to the main strength of the pristine GB, thus increase the strength of the pristine GB. For the pristine Σ11(332) GB in Fig. 11(b3), there is an obvious low charge density region between Al (2) and Al (2') atoms, but the segregation of Cu at [002] plane of the GB increases the charge density of this region significantly, as shown in Fig. 11(b4). The charge density of the new bonds Cu-Al (2) and Cu-Al (2') (0.04 e/bohr$^3$) is higher than that of Al (1)-Al (1') (0.038 e/bohr$^3$), indicating a strengthening effect of Cu segregation at Σ11(332) GB.

**3.5 Density of states analysis**

To further understand the characteristics of the chemical bond of Mg and Cu with Al atoms in different GBs, the partial density of state (PDOS) of Mg, Cu and their surrounding Al atom has been calculated. The Σ5(210) GB and Σ11(332) GB are chosen as examples for illustration to analyze the bonding characteristics. Fig. 12(a) shows that the contribution of s and p electron of the Al (4) and Al (4') atoms in Σ5(210) GB is almost the same. The hybridized peaks implying a strong electronic interaction happens between the two atoms, which agrees with the result acquired from the charge density analysis. Also, as shown in Fig. 12(b), the PDOS of Al (1) and Al (1') atom at the pristine Σ11(332) GB have a similar electron distribution on s and p orbit, indicating there is a strong chemical bond formed between the two atoms. The whole shape of the PDOS distribution of Al (4)-Al(4') and Al(1)-Al(1') is similar to the distribution of that of bulk Al, meaning that bond in the GB have similar characteristic to the bond in Al bulk. The formants number of Al (1)-Al (1') in Σ11(332) GB is less than that of Al (4)-Al (4') in Σ5(210) GB, which is consistence with the results of the charge density analysis and can explain the higher strength of Σ5(210) GB (10.7 GPa) than that of Σ11(332) GB (10.12 GPa).

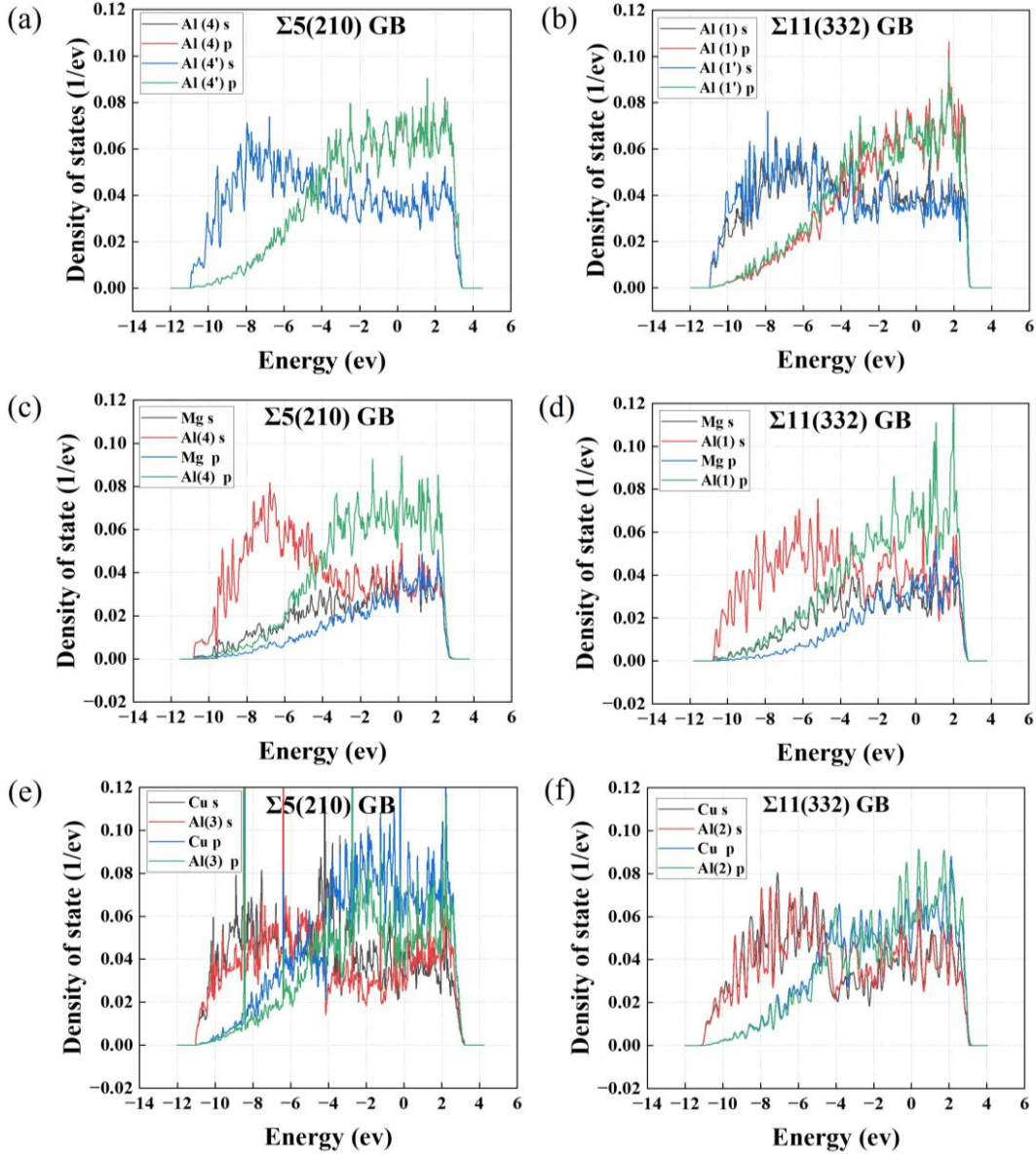

**Figure 12.** (a) PDOS(Partial density of state) of Al atom at Al bulk and Al atom at Al Σ5(210) GB, (b) PDOS of Al(3) and (-3) at pristine Al Σ5(210) GB, (c) PDOS of Al(1) and (-1) atom at pristine Al Σ9(221) GB, (d) PDOS of Mg atom and the Al(3) atom at Al Σ5(210) GB, (e) PDOS of Cu atom and the closest Al(2) atom at Al Σ5(210) GB, (f) PDOS of Mg atom and the Al(1) atom at Al Σ9(221) GB. (h)PDOS of Cu atom and the Al(2) atom at Al Σ5(221) GB.

Fig. 12(c) and (d) shows the s and p electron distribution of Mg and the nearest Al atom at Σ5(210) and Σ11(332) GB. The Mg atom has the same s electron with Al atom, but it was found that the total density of s electron of Mg is weaker than that of Al(4) and Al (1) at Σ5(210) GB and Al (1) at Σ11(332), indicating there is charge transfer between them, which caused the charge depletion around the GB. It can be found that fewer formants appeared in the s and p electron between the solute Mg and Al(4) atoms than that of Al (4) and Al (4') in Σ5(210) GB. A similar situation was found in Mg atom and Al (1) atom at Σ11(332) GB shown in Fig. 12(d), implying there are few electronic interactions between Al and Mg atoms. Although the results are not shown here, the segregation of Mg atom has a similar electronic effect at Σ5(310) and Σ9(221) GBs. That is, Mg solute does not form strong chemical bonds with Al atoms at all GBs. The weak Mg-Al bonds eventually lead to the weakening of the theoretical strength of Al GBs with the segregation of the Mg atom. Fig. 12(e) shows the s and p electron distribution of Cu atoms and the closest Al (3) atoms at Σ5(210) GB. It can be found that there are obvious formants on the s electron of Cu and Al (3) at -8.43ev and the p electron at -2.75ev, 2.25ev. Because of the similar distribution of PDOS, there are also many small formants on the s and p orbits at other values, which means that the solute Cu atom have strong electronic interaction with surrounding Al atoms. According to the PDOS calculation, the chemical bonds between Cu and Al (3) have similar characteristics to that of Al (4)-Al (4') at Σ5(210) GB. Likely, as shown in Fig. 12(f), the PDOS distribution of the Cu atom has a similar shape to that of the Al (2) atom at Σ11(332) GB, and there are many formants in s and p electron between them. It can be concluded that the Cu atom forms a strong chemical bond with the surrounding Al atoms when Cu segregates at Al GBs. Therefore, the strengthening of the theoretical strength of Al GBs with the segregation of Cu atoms can be ascribed to the strong electronic interaction between Cu and Al atoms.

## 4. Conclusions

First-principles calculations were carried out to investigate the effect of Mg and Cu segregation on the energy and mechanical properties of different Al GBs. The GB energy, segregation energy, and theoretical strength of GBs are calculated. According to the charge density analysis and density of states analysis, the relationship between Mg, Cu segregation, and the theoretical strength of GBs are explained. The main results are concluded as follows:

(1) By calculating the segregation energy of Mg and Cu solute atoms at different positions of GBs, the segregation tendency and segregation position are determined. The results show that both Mg and Cu atoms have a strong segregation tendency at the four studied Al GBs. Cu tends to form interstitial segregation, while Mg tends to reside in the GBs as substitutional segregation.

(2) Segregation of Mg and Cu solute atoms can significantly change the energy

characteristics of GBs. GB energy results show that both Cu and Mg can reduce GB energy, and the GB energy continues to decrease with the increase of the segregation concentration. The energy reduction effect of Cu on GBs is better than that of Mg solute, and Cu segregation can reduce the GB energy to negative values. According to the analysis of Bader charge, Cu has a stronger charge transfer effect with Al than that of Mg, indicating a stronger electronic interaction, which leads to a higher tendency to segregate at Al GBs than Mg.

(3) By performing the first-principles tensile tests, the effect of Mg and Cu solute atoms on the theoretical strength of GBs was obtained. The results show that the segregation of Mg has a negative effect on the strength of GBs, and the GB strength decreases with the increase of the Mg concentration, while the segregation of Cu enhances the GB strength, and the GB strength increases with the Cu concentration. Compared to the other two GBs, $\Sigma 5(210)$ GB and $\Sigma 9(221)$ GB showed a more significant effect of strengthening or weakening due to the solute atom segregation.

(4) According to the charge density analysis and density of states analysis, the segregation of Mg caused charge depletion and structure expansion at the Al GB plane, which lead to the decrease of GB strength. The segregation of Cu increases the charge density of GBs and forms new bonds with the surrounding Al atoms, which increase the strength of GBs.

## Declaration of Competing Interest

The authors declare no Competing Financial or No-Financial Interests.

## Acknowledgment

This work was supported by the National Natural Science Foundation of China (52071034; 52130107), the National Key Research and Development Program of China (2020YFA0405900), and the Fundamental Research Funds for the Central Universities (2022CDJKYJHJJ005).

*Supplementary Materials*

# Effect of solute atoms segregation on Al grain boundary properties by First-principles study

Xuan Zhang [a], Liang Zhang [a,b,*], Zhihui Zhang [a], Yasushi Shibuta [c], Xiaoxu Huang [a,b]

[a] *International Joint Laboratory for Light Alloys (MOE), College of Materials Science and Engineering, Chongqing University, Chongqing 400044, China*

[b] *Shenyang National Laboratory for Materials Science, Chongqing University, Chongqing 400044, China*

[c] *Department of Materials Engineering, The University of Tokyo, Bunkyo-ku, Tokyo 113-8656, Japan*

[*]Corresponding author. Email: liangz@cqu.edu.cn (L.Z.)

**Table S1.** Separation energy and tensile strength of Al Σ5(210) GB with different concentration (ML) of Mg and Cu. The results from RGS without relaxation and RGS with relaxation approaches are included.

| System | Separation energy (J/m$^2$) | | Tensile strength (GPa) | |
|---|---|---|---|---|
| | relaxation | un-relaxation | relaxation | un-relaxation |
| Bulk (210) plane | 2.07 | 2.16 | 9.69 | 11.93 |
| Al Σ5(210) GB | 1.57 | 1.75 | 8.68 | 10.71 |
| Al GB+0.25ML-Cu | 1.64 | 1.85 | 9.25 | 11.38 |
| Al GB+0.50ML-Cu | 1.71 | 1.95 | 9.30 | 11.86 |
| Al GB+0.75ML-Cu | 1.88 | 2.07 | 9.85 | 12.82 |
| Al GB+1.00ML-Cu | 2.04 | 2.15 | 10.45 | 13.45 |
| Al GB+0.25ML-Mg | 1.57 | 1.73 | 8.53 | 10.29 |
| Al GB+0.50ML-Mg | 1.59 | 1.72 | 8.35 | 10.01 |
| Al GB+0.75ML-Mg | 1.60 | 1.68 | 8.23 | 9.42 |
| Al GB+1.00ML-Mg | 1.62 | 1.67 | 7.84 | 8.95 |

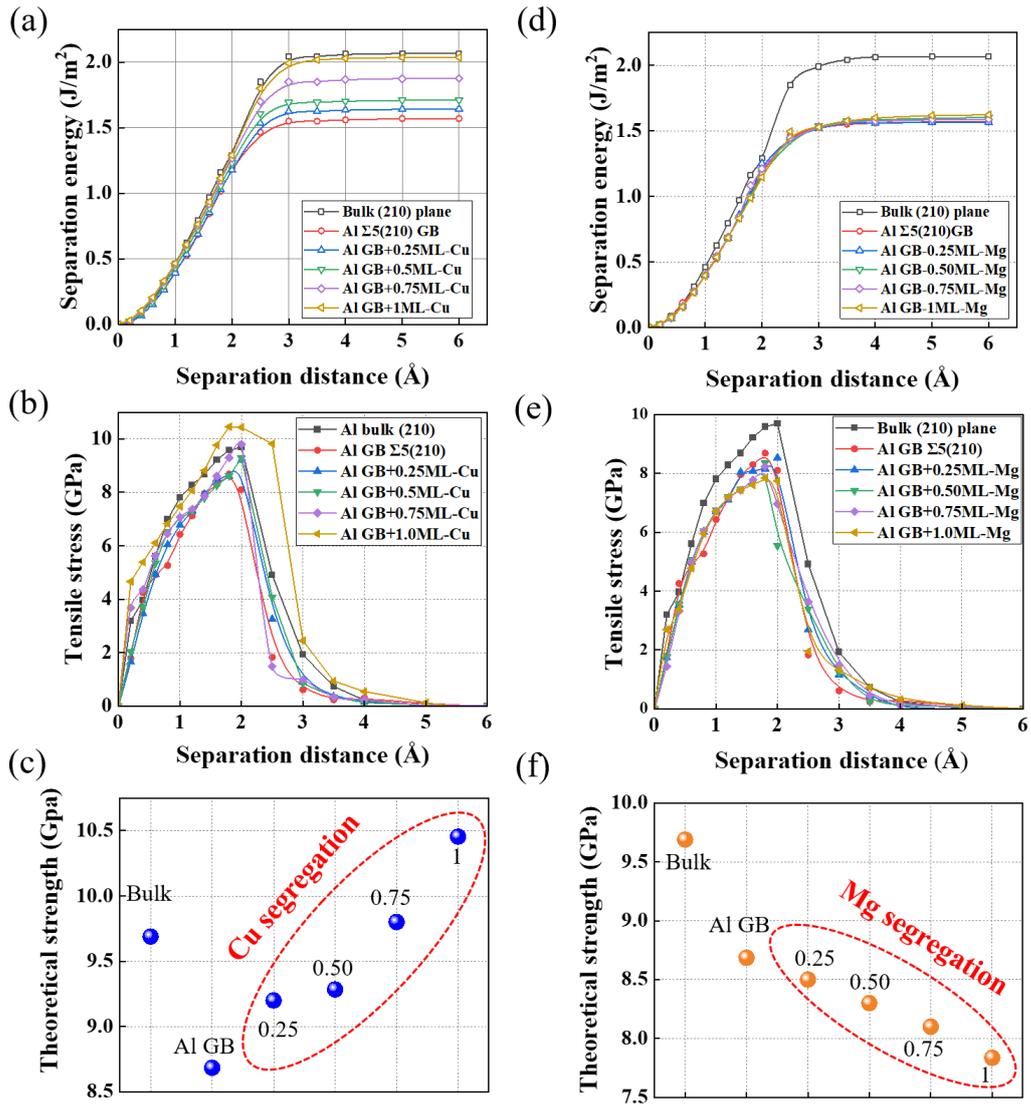

**Figure S1.** Separation energy and tensile stress of Σ5(210) GB with different segregation concentration of Cu, Mg solute atoms. (Calculations are performed using full atomic relaxation method)

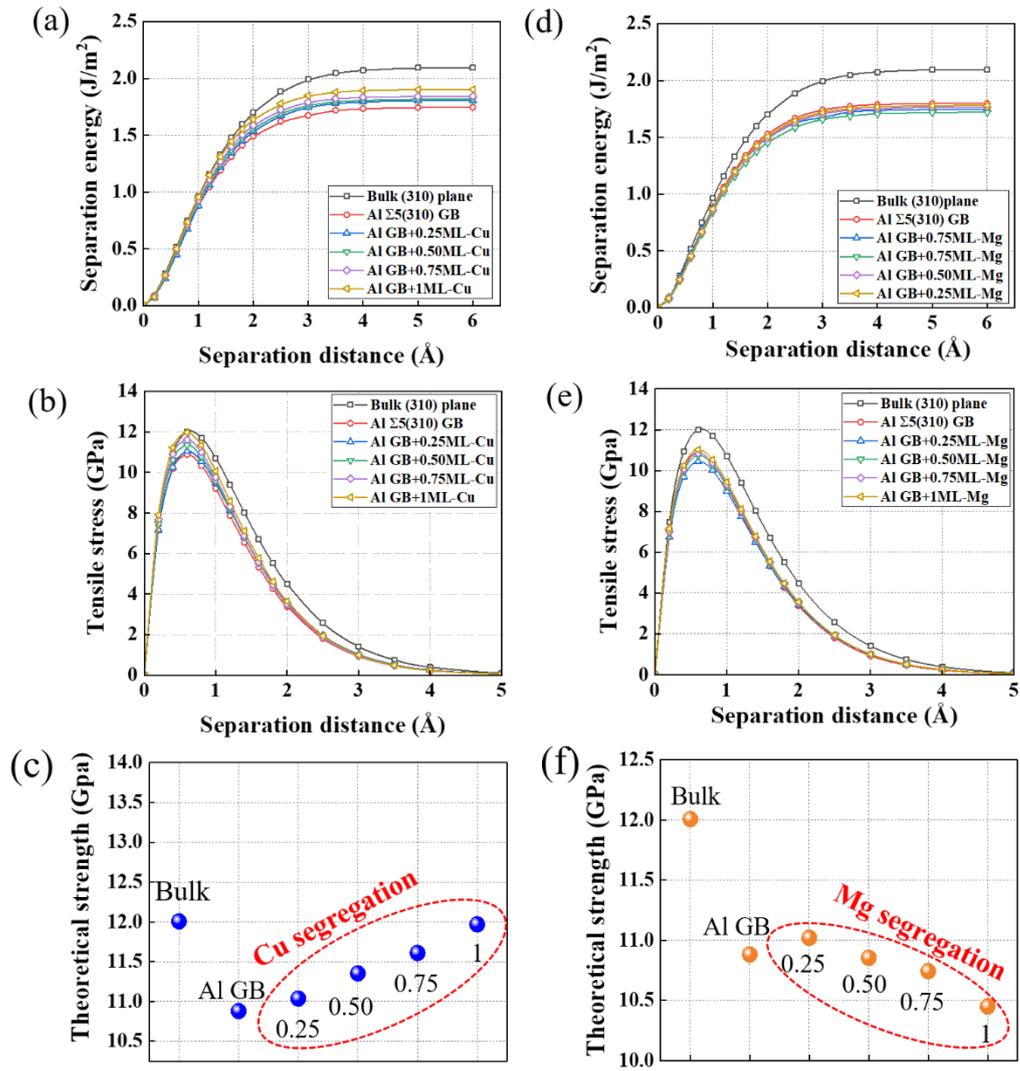

**Figure S2.** The separation energy and tensile stress as a function of the separation distance of Σ5(310) GB with different concentrations of Cu, and Mg segregation.

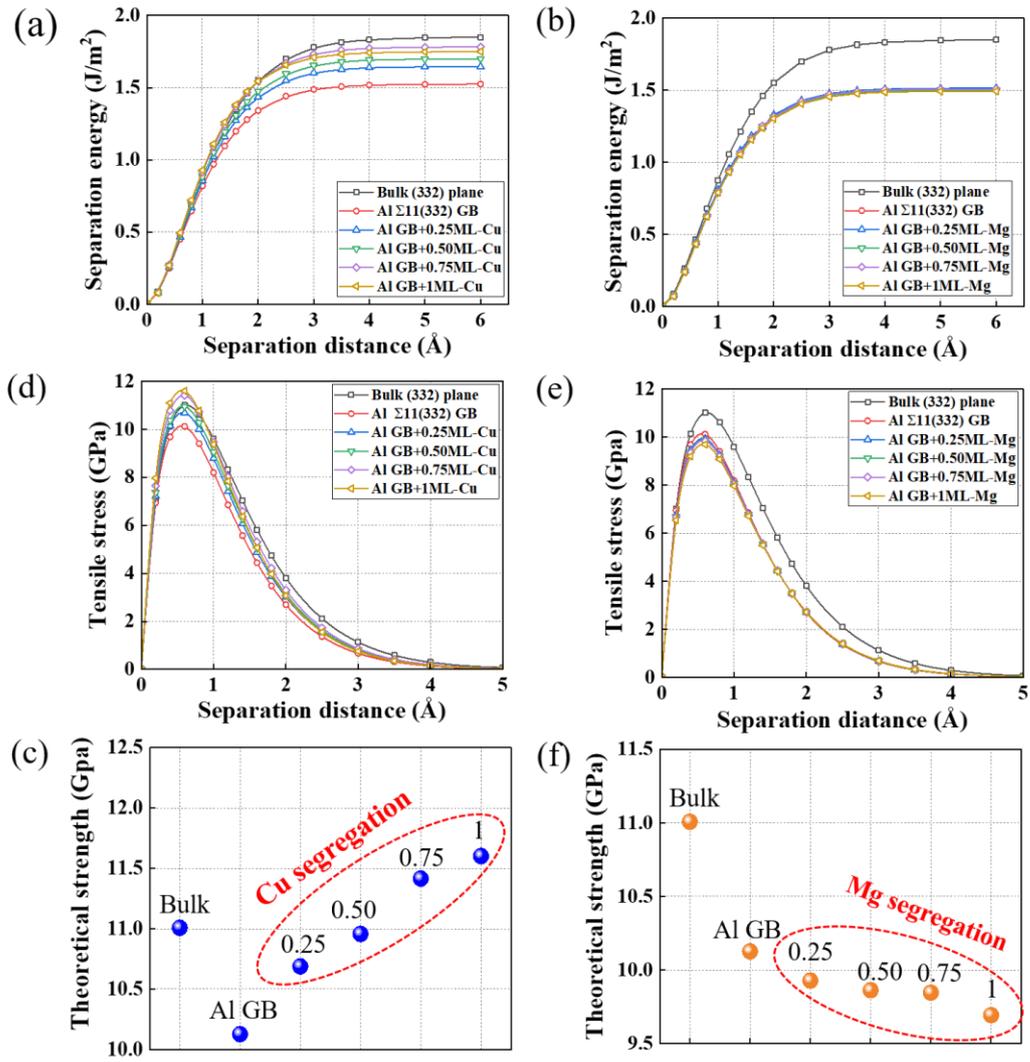

**Figure S3.** The separation energy and tensile stress as a function of the separation distance of Σ11(332) GB with different concentrations of Cu, and Mg segregation

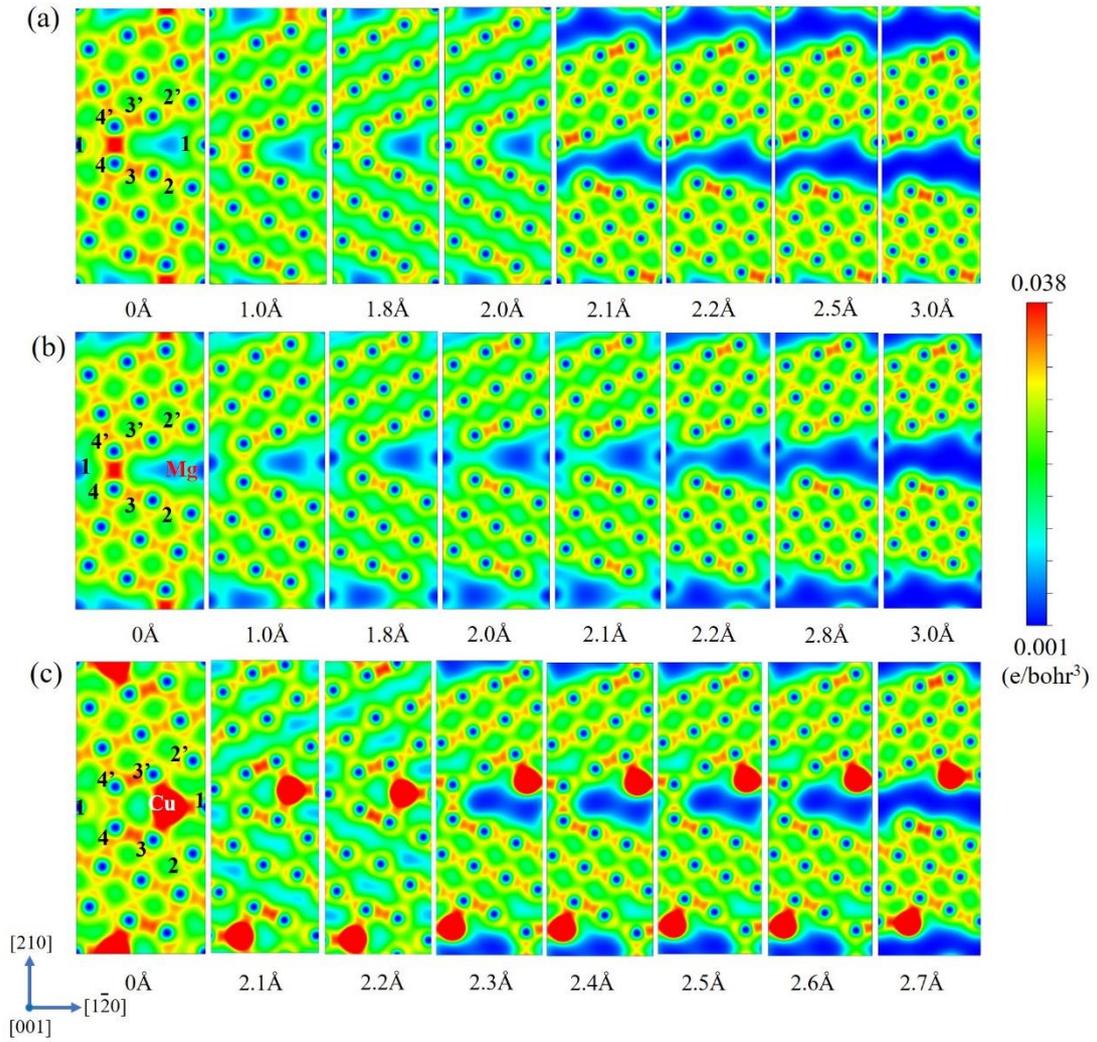

**Figure S4.** Charge density distributions of (a) the pristine Al Σ5(210) GB, (b) GB with Mg segregation, and (c) GB with Cu segregation at different separation distances. (Calculations are performed using full atomic relaxation method)